\documentstyle[12pt]{article}

\begin{document}

\setcounter{page}{1}

\pagestyle{plain} \vspace{1cm}
\begin{center}
\Large{\bf DGP Cosmology with a Non-Minimally Coupled Scalar Field on the Brane}\\
\small \vspace{1cm}
{\bf Kourosh Nozari$^{a,b}$}\\
\vspace{0.5cm} $^{a}$ {\it Centre for Particle Theory, Durham
University, South Road, Durham DH1 3LE, UK}\\
$^{b}${\it Department of Physics, Faculty of Basic Sciences,
University of Mazandaran,\\
P. O. Box 47416-1467,
Babolsar, IRAN\\
e-mail: knozari@umz.ac.ir}

\end{center}
\vspace{1.5cm}
\begin{abstract}
We construct a DGP inspired braneworld scenario where a scalar field
non-minimally coupled to the induced Ricci curvature is present on
the brane. First we investigate the status of gravitational
potential with non-minimal coupling and observational constraints on
this non-minimal model. Then we further deepen the idea of embedding
of FRW cosmology in this non-minimal setup. Cosmological
implications of this scenario are examined with details and the
quintessence and late-time expansion of the universe within this
framework are examined. Some observational constraints imposed on
this non-minimal scenario are studied and relation of this model
with dark radiation formalism is determined with details.

\end{abstract}
\vspace{2cm}
\newpage
\section{Introduction}
Based on light-curves analysis of several hundreds type Ia
supernovae [1,2], observations of the cosmic microwave background
radiation by the WMAP satellite [3] and other CMB-based experiments
[4,5], it has been revealed that our universe is currently in a
period of accelerated expansion. Some authors have attributed this
late-time expansion of the universe to an energy component referred
to as {\it dark energy}. The simplest example in this regard is the
cosmological constant itself which provides a model of dark energy.
However, it is unfavorable since it requires a huge amount of
fine-tuning [6]. Phantom fields [7], quintessence [8] and
modification of gravitational theory itself [9,10] are other
attempts to explain this late time expansion of the universe. In the
spirit of the modified gravitational theory, Carroll {\it et al}
have proposed  $R^{-1}$ modification of the usual Einstein-Hilbert
action [11]. It was then shown that this term could give rise to
accelerating solutions of the field equations without dark energy.

On the other hand, theories of extra spatial dimensions, in which
the observed universe is realized as a brane embedded in a higher
dimensional spacetime, have attracted a lot of attention in the last
few years. In this framework, ordinary matters are trapped on the
brane but gravitation propagates through the entire spacetime
[9,12,13]. The cosmological evolution on the brane is given by an
effective Friedmann equation that incorporates the effects of the
bulk in a non-trivial manner [14]. From a cosmological view point,
the importance of brane models lies, among other things, in the fact
that they can provide an alternative scenario to explain the
late-time accelerating expansion of the universe.

Theories with extra dimensions usually yield correct Newtonian limit
at large distances since the gravitational field is quenched on
sub-millimeter transverse scales. This quenching appears either due
to finite extension of the transverse dimensions [12,15] or due to
sub-millimeter transverse curvature scales induced by negative
cosmological constant [13,16-19]. A common feature of these type of
models is that they predict deviations from the usual 4-dimensional
gravity at short distances. The model proposed by Dvali, Gabadadze
and Porrati (DGP) [9] is different in this regard since it predicts
deviations from the standard 4-dimensional gravity even over large
distances. In this scenario, the transition between four and
higher-dimensional gravitational potentials arises due to the
presence of both the brane and bulk Einstein terms in the action. In
this framework, existence of a higher dimensional embedding space
allows for the existence of bulk or brane matter which can certainly
influence the cosmological evolution on the brane. Even if there is
no 4-dimensional Einstein-Hilbert term in the classical theory, such
a term should be induced by loop-corrections from matter fields
[20,21]. Generally one can consider the effect of an induced gravity
term as a quantum correction in any brane-world scenario.\\
A particular form of bulk or brane matter is a scalar field. Scalar
fields play an important role both in models of the early universe
and late-time acceleration. These scalar fields provide a simple
dynamical model for matter fields in a brane-world model. In the
context of induced gravity corrections, it is then natural to
consider a non-minimal coupling of the scalar field to the intrinsic
(Ricci) curvature on the brane that is a function of the field. The
resulting theory can be thought of as a generalization of
Brans-Dicke type scalar-tensor gravity in a brane-world context. In
contrast to common belief, the introduction of non-minimal coupling
(NMC) is not a matter of taste; NMC is instead forced upon us in
many situations of physical and cosmological interest. There are
many compelling reasons to include an explicit non-minimal coupling
in the action. For instance, NMC arises at the quantum level when
quantum corrections to the scalar field theory are considered. Even
if for the classical, unperturbed theory this NMC vanishes, it is
necessary for the renormalizability of the scalar field theory in
curved space. In most theories used to describe inflationary
scenarios, it turns out that a non-vanishing value of the coupling
constant cannot be avoided. In general relativity, and in all other
metric theories of gravity in which the scalar field is not part of
the gravitational sector, the coupling constant necessarily assumes
the value of \, $\frac{1}{6}$\,. The study of the asymptotically
free theories in an external gravitational field with a Gauss-Bonnet
term shows a scale dependent coupling parameter. Asymptotically free
grand unified theories have a non-minimal coupling depending on a
renormalization group parameter that converges to the value of
$\frac{1}{6}$ or to any other initial conditions depending on the
gauge group and on the matter content of the theory. An exact
renormalization group study of the $\lambda\phi^4$ theory shows that
NMC$=\frac{1}{6}$ is a stable infrared fixed point. Also in the
large $N$ limit of the Nambu-Jona-Lasinio model, we have
NMC$=\frac{1}{6}$. In the $O(N)$- symmetric model with $V
=\lambda\phi^4$, NMC  is generally nonzero and depends on the
coupling constants of the individual bosonic components. Higgs
fields in the standard model have NMC$=0$ or $\frac{1}{6}$. Only a
few investigations produce zero value(for a more complete discussion
of these issues we refer to papers by V. Faraoni, specially Ref.
[22] and references therein). In view of the above results, it is
then natural to incorporate an explicit NMC between scalar field and
Ricci scalar in the inflationary paradigm and in quintessence
models. In particular it is interesting to see the effect of this
NMC in a DGP-inspired braneworld cosmology.

There are several studies focusing on the braneworld models with
scalar field [22-31]. Some of these studies are concentrated on the
bulk scalar field minimally [23-25] or non-minimally[26-28] coupled
to the bulk Ricci scalar. Some other authors have studied the
minimally [29,30] or non-minimally [22,31] coupled scalar field to
the induced Ricci scalar on the brane. However, none of these
studies have considered consequences of embedding of FRW cosmology
with non-minimally coupled brane-scalar field into DGP scenario
explicitly. Specifically, none of these studies obtained five
dimensional metric components and full dynamics of the braneworld
within this scenario trivially. In addition, there are very limited
literature on late time behavior and quintessence with non-minimally
coupled scalar field and even these studies have not performed
explicit and detailed calculation of the late-time dynamics.\\
In this paper, in the spirit of DGP inspired gravity, we study the
effect of an induced gravity term which is an arbitrary function of
a scalar field on the brane. We present four-dimensional equations
on a DGP brane with a scalar field non-minimally coupled to the
induced Ricci curvature, embedded in a five-dimensional Minkowski
bulk. This is an extension to a braneworld context of scalar-tensor
(Brans-Dicke) gravity. We show that our model allows for an
embedding of the standard Friedmann cosmology in the sense that the
cosmological evolution of the background metric on the brane can
entirely be described by the standard Friedmann equation plus energy
non-conservation on the brane. In this framework we explore the
relation between our formalism and the so-called dark radiation
formalism. We study cosmological implications of both minimal and
non-minimal extension of our model. In minimal case, motivated by
modified theories of gravity, the potential describing minimally
coupled scalar field is taken to be that with $R^{-1}$ term added to
the usual Einstein-Hilbert action[11]. In non-minimal case however,
we concentrate mainly on the potential of the type
$V(\phi)=\lambda{\phi^n}$. We study the weak field limit of our
model and show that the mass density of ordinary matter on the brane
should be modified by the addition of the effective mass density
attributed to the non-minimally coupled scalar field on the brane.
Considering the case of FRW brane, we obtain the evolution of the
metric and scalar field by solving the field equations in the limit
of small curvature. Our solutions for minimal case predict a
power-law acceleration on the brane supporting observed late-time
acceleration. For non-minimal case (by adapting a simple ansatz) we
show that by a suitable choice of non-minimal coupling and scalar
field potential one can achieve accelerated expansion in some
special cases. We study quintessence model with non-minimally
coupled scalar field on the brane and discuss some observational
constraints imposed on the value of the non-minimal coupling using
supernova data.\\
We use a prime for differentiation with respect to fifth coordinate
except for two special cases: $\alpha'\equiv\frac{d\alpha}{d\phi}$
and $V'\equiv\frac{d V}{d\phi}$. An overdot denotes differentiation
with respect to the comoving time, $t$.

\section{Induced Gravity with Non-Minimally Coupled Brane-Scalar
Field}

The action of the DGP scenario in the presence of a non-minimally
coupled scalar field on the brane can be written as follows
\begin{equation}
S=\int d^{5}x\frac{m^{3}_{4}}{2}\sqrt{-g}{\cal R}+\Bigg[\int
d^{4}x\sqrt{-q}\bigg(\frac{m_{3}^{2}}{2}\alpha(\phi)
R[q]-\frac{1}{2} q^{\mu\nu} \nabla_{\mu}\phi\nabla_{\nu}\phi
-V(\phi) + m^{3}_{4}\overline{K}+ {\cal{L}}_{m}\bigg)\Bigg]_{y=0},
\end{equation}
where we have included a general non-minimal coupling $\alpha(\phi)$
\,in the brane part of the action( for an interesting discussion on
the possible schemes to incorporate NMC in the formulation of
scalar-tensor gravity see [22,35]).\, $y$ is coordinate of the fifth
dimension and we assume brane is located at $y=0$.\, $g_{AB}$ is
five dimensional bulk metric with Ricci scalar ${\cal{R}}$, while
$q_{\mu\nu}$ is induced metric on the brane with induced Ricci
scalar $R$.\, $g_{AB}$ and $q_{\mu\nu}$ are related via
$q_{\mu\nu}={\delta_{\mu}}^{A}{\delta_{\nu}}^{B}g_{AB}$.\,
$\overline{K}$ is trace of the mean extrinsic curvature of the brane
defined as
\begin{equation}
\overline{K}_{\mu\nu}=\frac{1}{2}\,\,\lim_{\epsilon\rightarrow
0}\bigg(\Big[K_{\mu\nu}\Big]_{y=-\epsilon}+
\Big[K_{\mu\nu}\Big]_{y=+\epsilon}\bigg),
\end{equation}
and corresponding term in the action is York-Gibbons-Hawking
term[32] (see also [20]). The ordinary matter part of the action is
shown by Lagrangian ${\cal{L}}_{m}\equiv
{\cal{L}}_{m}(q_{\mu\nu},\psi)$ where $\psi$ is matter field and
corresponding energy-momentum tensor is
\begin{equation}
T_{\mu\nu}=-2\frac{\delta{\cal{L}}_{m}}{\delta
q^{\mu\nu}}+q_{\mu\nu}{\cal{L}}_{m}.
\end{equation}
The pure scalar field Lagrangian,\, ${\cal{L}}_{\phi}=-\frac{1}{2}
q^{\mu\nu} \nabla_{\mu}\phi\nabla_{\nu}\phi -V(\phi)$,\,\,  yields
the following energy-momentum tensor
\begin{equation}
 \tau_{\mu\nu}=\nabla_\mu\phi\nabla_\nu\phi-\frac{1}{2}q_{\mu\nu}(\nabla\phi)^2
-q_{\mu\nu}V(\phi)
\end{equation}
The Bulk-brane Einstein's equations calculated from action (1) are
given by
$$m^{3}_{4}\left({\cal R}_{AB}-\frac{1}{2}g_{AB}{\cal
R}\right)+$$
\begin{equation}
m^{2}_{3}{\delta_{A}}^{\mu}{\delta_{B}}^{\nu}\bigg[\alpha(\phi)\left(R_{\mu\nu}-
\frac{1}{2}q_{\mu\nu}R\right)-\nabla_{\mu}\nabla_{\nu}\alpha(\phi)+q_{\mu\nu}\Box^{(4)}\alpha(\phi)\bigg]\delta(y)
={\delta_{A}}^{\mu}{\delta_{B}}^{\nu}\Upsilon_{\mu\nu}\delta(y),
\end{equation}
where $\Box^{(4)}$ is 4-dimensional(brane) d'Alembertian and
$\Upsilon_{\mu\nu}=T_{\mu\nu}+\tau_{\mu\nu}$\,. This relation can be
rewritten as follows
\begin{equation}
m^{3}_{4}\left({\cal R}_{AB}-\frac{1}{2}g_{AB}{\cal R}\right)+
m^{2}_{3}\alpha(\phi){\delta_{A}}^{\mu}{\delta_{B}}^{\nu}\left(R_{\mu\nu}-
\frac{1}{2}q_{\mu\nu}R\right)\delta(y)=
{\delta_{A}}^{\mu}{\delta_{B}}^{\nu}{\cal{T}}_{\mu\nu}\delta(y)
\end{equation}
where ${\cal{T}}_{\mu\nu}$ is total energy-momentum on the brane
defined as follows
\begin{equation}
{\cal{T}}_{\mu\nu}=m^{2}_{3}\nabla_{\mu}\nabla_{\nu}\alpha(\phi)-m^{2}_{3}
q_{\mu\nu}\Box^{(4)}\alpha(\phi)+\Upsilon_{\mu\nu},
\end{equation}
From (6) we find
\begin{equation}
G_{AB}={\cal R}_{AB}-\frac{1}{2}g_{AB}{\cal R}=0
\end{equation}
and
\begin{equation}
G_{\mu\nu}=\left(R_{\mu\nu}-
\frac{1}{2}q_{\mu\nu}R\right)=\frac{{\cal
T}_{\mu\nu}}{m^{2}_{3}\alpha(\phi)}
\end{equation}
for bulk and brane respectively. The corresponding junction
conditions relating the extrinsic curvature to the energy-momentum
tensor of the brane, have the following form
\begin{equation}
\lim_{\epsilon\rightarrow+0}\Big[K_{\mu\nu}\Big]^{y=+\epsilon}_{y=-\epsilon}
=\frac{1}{m_{4}^{3}}\bigg[{\cal{T}}_{\mu\nu}-\frac{1}{3}q_{\mu\nu}q^{\alpha\beta}
{\cal {T}}_{\alpha\beta}\bigg]_{y=0}
-\frac{m^{2}_{3}\alpha(\phi)}{m^{3}_{4}}\bigg[R_{\mu\nu}-
\frac{1}{6}q_{\mu\nu}q^{\alpha\beta}R_{\alpha\beta}\bigg]_{y=0}.
\end{equation}

We set\, $g_{AB}=\eta_{AB}+h_{AB}$ \, where $h_ {AB}$ are small
perturbations to investigate the weak field limit of the scenario.
Within the Gaussian normal coordinate, we impose the following
harmonic gauge on the longitudinal coordinates[9]
\begin{equation}
\partial_\alpha h^\alpha{}_\mu+\partial_{y} h_{y\mu}
=\frac{1}{2}\partial_\mu\left(h^\alpha{}_\alpha +h_{yy}\right).
\end{equation}
This will led us to a decoupled equation for the gravitational
potential of a static mass distribution.  The transverse equations
in the gauge (11) give $h_{y\mu}=0$, $h_{yy}=h^\alpha{}_\alpha$. The
remaining equations take the following form
\begin{equation}
m^{3}_{4}(\partial_{\alpha}\partial^{\alpha}+\partial_{y}^{2})h_{\mu\nu}+m^{2}_{3}\alpha(\phi)\left(\partial_{\alpha}\partial^{\alpha}
h_{\mu\nu}-\partial_{\mu}\partial_{\nu}{h_{\alpha}}^{\alpha}\right)\delta(y)=
-2\delta(y)\left[{\cal
{T}}_{\mu\nu}-\frac{1}{3}\eta_{\mu\nu}\eta^{\alpha\beta}{\cal
{T}}_{\alpha\beta}\right].
\end{equation}
We suppose that non-minimally coupled scalar field has an effective
mass $M_{\phi}$. The gravitational potential of mass densities
$\rho_{\psi}(\vec{r})=M_{\psi}\delta(\vec{r})$ and
$\rho_{\phi}(\vec{r})=M_{\phi}\delta(\vec{r})$ on the brane
satisfies the following equation
\begin{equation}
m^{3}_{4}\left(\partial_{\alpha}\partial^{\alpha}+\partial^{2}_{y}\right)U(\vec{r},y)+
m^{2}_{3}\alpha(\phi)\delta(y)\partial_{\alpha}\partial^{\alpha}U(\vec{r},y)=\frac{2}{3}
(M_{\psi}+M_{\phi})\delta(\vec{r})\delta(y).
\end{equation}
This equation shows that in the presence of non-minimally coupled
scalar field, the mass in standard DGP framework should be modified
by the addition of the mass of the non-minimally coupled scalar
field. Using the following Fourier ansatz
\begin{equation}
U(\vec{r},y)=\frac{1}{(2\pi)^4} \int d^3\vec{p}\int dp_{y}\,
U(\vec{p},p_{y}) \exp\Big[i(\vec{p}\cdot\vec{r}+p_{y} y)\Big]
\end{equation}
in equation (13), we find
\begin{equation}
m_4^3(\vec{p}^2+p_{y}^2)U(\vec{p},p_{y})+\frac{m_3^2\alpha(\phi)}{2\pi}
 \vec{p}^2\int dp_{y}'\,U(\vec{p},p_{y}')=-\frac{2}{3}(M_{\psi}+M_{\phi}).
\end{equation}
This integral equation has the following solution
\begin{equation}
U(\vec{p},p_{y})=-\frac{4}{3}\frac{M_{\psi}+M_{\phi}}{(\vec{p}^2+p_{y}^2)
\Big(2m_{4}^3+m_{3}^2\alpha(\phi)|\vec{p}|\Big)}.
\end{equation}
The resulting potential on the brane is
\begin{equation}
U(\vec{r})=-\bigg(\frac{M_{\psi}+M_{\phi}}{6\pi m_3^2\alpha(\phi)
r}\bigg)\bigg[ \cos(\xi_{\alpha} r)- \frac{2}{\pi}\cos(\xi_{\alpha}
r)\, \mbox{Si}(\xi_{\alpha} r) +\frac{2}{\pi}\sin(\xi_{\alpha} r)
\,\mbox{Ci}(\xi_{\alpha} r) \bigg],
\end{equation}
where $\xi_{\alpha}=\frac{2m_4^3}{m_3^2\alpha(\phi)}$\, and the {\it
sine}  and {\it cosine} integrals are defined as follows
$$\mbox{Si}(x)=\int_0^xd\omega\,\frac{\sin\omega}{\omega},\quad\quad\quad\quad
\mbox{Ci}(x)=-\int_x^\infty d\omega\,\frac{\cos\omega}{\omega}.$$
Now there is a modified transition scale
\begin{equation}
{\xi_{\alpha}}^{-1}\equiv\ell_{\alpha}=\frac{m_3^2\alpha(\phi)}{2m_4^3}=\alpha(\phi)\ell_{DGP}
\end{equation}
between four and five-dimensional behavior of the gravitational
potential in this scenario:
\begin{equation}
r\ll\ell_{\alpha}:\quad\quad
U(\vec{r})=-\frac{M_{\psi}+M_{\phi}}{6\pi m_3^2\alpha(\phi) r}
\bigg[1+\Big(\gamma-\frac{2}{\pi}\Big)\frac{r}{\ell_{\alpha}}
+\frac{r}{\ell_{\alpha}}\ln\Big(\frac{r}{\ell_{\alpha}}\Big)
+{\cal{O}}\Big(\frac{r^2}{\ell_{\alpha}^2}\Big)\bigg],
\end{equation}
and
\begin{equation}
r\gg\ell_{\alpha}:\quad\quad\quad\quad
U(\vec{r})=-\frac{M_{\psi}+M_{\phi}}{6\pi^2 m_4^3 r^2}
\bigg[1-2\frac{\ell_{\alpha}^2}{r^2}
+{\cal{O}}\Big(\frac{\ell_{\alpha}^4}{r^4}\Big)\bigg],
\end{equation}
where \, $\gamma=0.577$\, is Euler's constant. Therefore, existence
of non-minimally coupled scalar field on the brane has the following
consequences: the mass density of ordinary matter on the brane
should be modified by the addition of the mass density attributed to
the scalar field on the brane and the DGP transition scale between
four and five dimensional behavior of gravitational potential now is
explicitly dependent on the strength of non-minimal coupling. If
$\alpha(\phi)$ varies slightly from point to point on the brane, it
can be interpreted as a spacetime dependent Newton's constant.  The
dynamics that control this variation are determined by the following
equation
\begin{equation}
\nabla^{\mu}\phi\nabla_{\mu}\phi-\frac{dV}{d\phi}+\frac{m_{3}^{2}}{2}\Big(\frac{d\alpha(\phi)}{d\phi}\Big)R=0.
\end{equation}
Note that in the real world we don't want $\alpha(\phi)$ to vary too
much since it will have observable effects in classic experimental
tests of general relativity and also in cosmological tests such as
primordial nucleosynthesis[33]. This can be ensured either by
choosing a large mass for scalar field $\phi$ or choosing
$\alpha(\phi)$ so that large changes in $\phi$ give rise to
relatively small changes in Newton's constant. So, when
$\alpha(\phi)$ varies in DGP brane from point to point, the
crossover scale will change and is no longer a constant. This
feature would change previous picture of crossover scale in DGP
scenario and my change some arguments on phenomenology of this
scenario[34]. Note that we can define a modified four-dimensional
Planck mass as $m_{3}^{(\alpha)}=\sqrt{\alpha(\phi)}m_{3}$ and
therefore the effect of non-minimal coupling can be attributed to
the modification of four dimensional Planck mass. This is equivalent
to modification of four dimensional Newton's constant[22]. If we use
the reduced Planck mass for $m_3$, gravitational potential for small
$r$ limit will differ from the ordinary four-dimensional potential
by a factor $\frac{4}{3}\alpha^{-1}$. In fact the coupling of the
masses on the brane to the induced Ricci tensor on the brane is
modified by this factor. Existence of this extra factor is in
agreement with the tensorial structure of the graviton propagator
due to additional helicity state of the five-dimensional graviton
[9,34]. Figure $1$ shows the shape of the DGP potential for
$r\ll\ell_{\alpha}$ (equation (19)) with some arbitrary values of
the NMC. As this figure shows, for large negative values of NMC, it
is possible to have repulsive gravitational potential. For positive
values of NMC the DGP potential for $r\ll\ell_{\alpha}$ is always
attractive. Figure $2$ shows the DGP potential for $r
\gg\ell_{\alpha}$ (equation (20)).\\

\begin{figure}[htp]
\begin{center}
\includegraphics{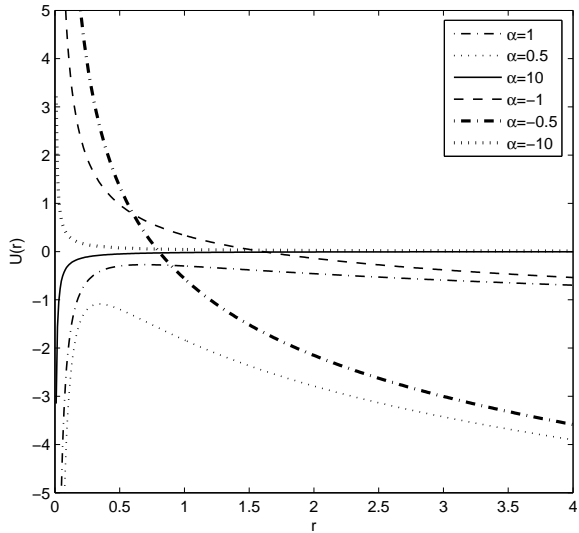}
\end{center}
\vspace{6 cm}
 \caption{\small { DGP potential for $r\ll\ell_{\alpha}$ and with some
 arbitrary values of NMC (equation (19)). }}
 \label{fig:1}
\end{figure}

\begin{figure}[htp]
\begin{center}
\includegraphics{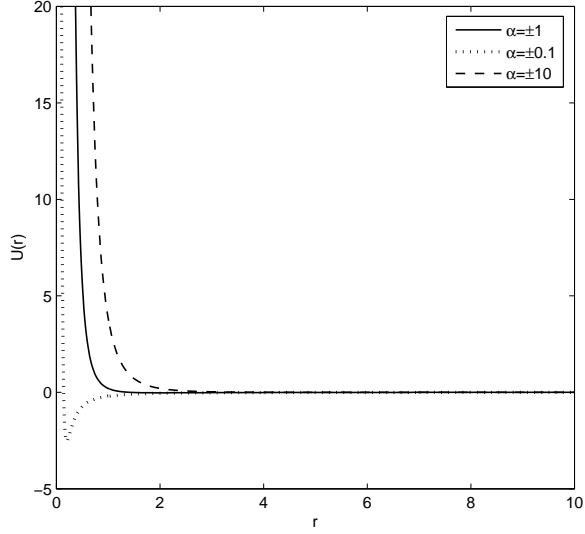}
\end{center}
\vspace{6 cm}
 \caption{\small { DGP potential for $r \gg \ell_{\alpha}$ and with some
 arbitrary values of NMC (equation (20)). }}
 \label{fig:1}
\end{figure}

Ignoring the tensor structure, the gravitational potential of a
source of mass $m$ is given by $U_{grav}\sim -\frac{G_{brane}m}{r}$
for $r\ll\ell_{\alpha}$ and $U_{grav}\sim -\frac{G_{bulk}m}{r^2}$
for $r \gg\ell_{\alpha}$ respectively. That is, the potential
exhibits four-dimensional behavior at short distances and
five-dimensional behavior (i.e., as if the brane were not there at
all) at large distances. In the absence of NMC, for the crossover
scale to be large, we need a substantial mismatch between
four-dimensional Planck scale (corresponding to the usual Newton's
constant, $G_{brane} = G$) and the fundamental, or bulk, Planck
scale $M_{4}$ [34]. However, with NMC $\alpha(\phi)$, evidently
$\ell_{\alpha}$ can be large with large values of $\alpha(\phi)$.
Therefore existence of NMC changes the mismatch condition between
four-dimensional Planck scale and the fundamental Planck scale. In
minimal case, the fundamental Planck scale $M_{4}$ has to be quite
small in order for the energy of gravity fluctuations to be
substantially smaller in the bulk versus on the brane, the energy of
the latter being controlled by $M_{3}=M_{P}$. This situation can be
mediate due to the presence of NMC. Note that when $M_{4}$ is small,
the corresponding Newton's constant in the bulk, $G_{bulk}$, is
large. So, for a given source mass $m$, gravity is much stronger in
the bulk. Earlier work using Supernova data implies that the
best-fit for density parameter of ordinary matter is given by
$\Omega^{0}_{M}=0.18^{+0.07}_{-0.06}$ [41,42]. We may introduce a
new effective dark energy component, $\Omega_{\ell_{\alpha}}$, where
$\Omega_{\ell_{\alpha}}=\frac{1}{\ell_{\alpha}H}$ to resort the
identity: $1 = \Omega_{M} + \Omega_{\ell_{\alpha}}$. Therefore we
find $\ell_{\alpha}=\Big(1.21^{+0.09}_{-0.09}\Big) H_{0}^{-1}$.
Assuming a flat universe, a more recent supernova data [43] suggests
a best-fit of $\Omega_{ M}= 0.21$ corresponding to a best-fit
crossover scale $\ell_{\alpha} = 1.26H_{0}^{-1}$. Since
$\ell_{\alpha}=\frac{m_3^2\alpha(\phi)}{2m_4^3}$ or
$\alpha(\phi)=\frac{2m_{4}^{3}\ell_{\alpha}}{m_3^2}$, for $M_{3}\sim
10^{18} GeV$ and $M_{4}\sim 10^{3} GeV$, we find $ \alpha(\phi)\sim
2.52 \times 10^{-30} H_{0}^{-1}$.

\section{Non-Minimal DGP Cosmology }
It has been shown that DGP model can account for the standard
Friedmann cosmology at any distance scale on the brane [20]. In our
proposed framework, we start with the following line element to
derive cosmological implications of our model,
\begin{equation}
ds^{2}=q_{\mu\nu}dx^{\mu}dx^{\nu}+b^{2}(y,t)dy^{2}=-n^{2}(y,t)dt^{2}+
a^{2}(y,t)\gamma_{ij}dx^{i}dx^{j}+b^{2}(y,t)dy^{2}.
\end{equation}
In this relation $\gamma_{ij}$ is a maximally symmetric
3-dimensional metric defined as
\begin{equation}
\gamma_{ij}=\delta_{ij}+k\frac{x_{i}x_{j}}{1-kr^{2}}
\end{equation}
where $k=-1,0,1$ parameterizes the spatial curvature and
$r^2=x_{i}x^{i}$. We assume that scalar field $\phi$ depends only on
the proper cosmic time of the brane. Choosing gauge $b^{2}(y,t)=1$
in Gaussian normal coordinates, the field equations in the bulk are
given by (8) with the following Einstein's tensor components
\begin{equation}
G_{00}=3n^{2} \Big(\frac{\dot{a}^2}{n^2a^2}-\frac{{a'}^2}{a^2}
-\frac{a''}{a}+\frac{k}{a^2}\Big),
\end{equation}
\begin{equation}
G_{ij}=\gamma_{ij}a^{2}\bigg[\Big(\frac{{a'}^2}{a^2}
-\frac{\dot{a}^2}{n^2a^2}-\frac{k}{a^2}\Big)
+2\Big(\frac{a''}{a}+\frac{n'a'}{na}
-\frac{\ddot{a}}{n^2a}+\frac{\dot{n}\dot{a}}{n^3a}+\frac{n''}{2n}\Big)\Bigg]
\end{equation}
\begin{equation}
G_{0y}=3\Big( \frac{n'}{n}\frac{\dot{a}}{a}-\frac{\dot{a}'}{a}
\Big),
\end{equation}
\begin{equation}
G_{yy}=3\Big(\frac{{a'}^2}{a^2}-\frac{\dot{a}^2}{n^2a^2}
-\frac{k}{a^2} +\frac{n'a'}{na}+\frac{\dot{n}\dot{a}}{n^3a}
-\frac{\ddot{a}}{n^2a}\Big).
\end{equation}
The field equations on the brane are given by the following
equations
\begin{equation}
G^{(3)}_{00}=3n^{2}\left(\frac{\dot{a}^{2}}{n^{2}a^{2}}+
\frac{k}{a^{2}}\right)=\frac{2}{m^{2}_{3}\alpha(\phi)}{\cal
{T}}_{00},
\end{equation}
\begin{equation}
G^{(3)}_{ij}=\gamma_{ij}\bigg[2\bigg(\frac{\dot{n}\dot{a}}{n^{3}a}-
\frac{\ddot{a}}{n^{2}a}\bigg)-\bigg(\frac{\dot{a}^{2}}{n^{2}a^{2}}+
\frac{k}{a^{2}}\bigg)\bigg]=\frac{2}{m^{2}_{3}\alpha(\phi)}{\cal
T}_{ij},
\end{equation}
and scalar field evolution equation
\begin{equation}
\ddot{\phi}+\bigg(3\frac{\dot a}{a}-\frac{\dot
n}{n}\bigg)\dot{\phi}+n^{2}\frac{dV}{d\phi}-\frac{m_{3}^2}{2}n^{2}\alpha'R[q]=0,
\end{equation}
where Ricci scalar on the brane is given by
\begin{equation}
R=3\frac{k}{a^2}+\frac{1}{n^{2}}\bigg[6\frac{\ddot{a}}{a}+6\Big(\frac{\dot{a}}{a}\Big)^{2}-6\frac{\dot{a}}{a}\frac{\dot{n}}{n}\bigg].
\end{equation}
The other important equation is the continuity equation on the
brane. Suppose that ordinary matter on the brane has an ideal fluid
form, $T_{\mu\nu}=(\rho+p)u_{\mu}u_{\nu}+pq_{\mu\nu}$. Since
$K_{tt}=nn'$ and $K_{rr}=-aa'$, equation (10) gives the following
matching conditions
\begin{equation}
\lim_{\epsilon\rightarrow+0}\left[\partial_{y}a\right]^{y=
+\epsilon}_{y=-\epsilon}(t)=\frac{m^{2}_{3}}{m^{3}_{4}}\Bigg[\alpha(\phi)
\bigg(\frac{\dot{a}^{2}}{n^{2}a}+ \frac{k}{a}\bigg)\Bigg]_{y=0}
-\Bigg[\frac{(\rho+\rho_{\phi})a}{3 m^{3}_{4}}\Bigg]_{y=0}.
\end{equation}
$$\lim_{\epsilon\rightarrow+0}\left[\partial_{y}n\right]^{y=
+\epsilon}_{y=-\epsilon}(t)=\frac{m^{2}_{3}}{m^{3}_{4}}(2n)\Bigg[\alpha(\phi)\Big(\frac{\ddot{a}}{n^{2}a}-\frac{\dot{a}^{2}}{2n^{2}a^2}-
\frac{\dot{n}\dot{a}}{n^{3}a}-\frac{k}{2a^{2}}\Big) \Bigg]_{y=0}+$$
\begin{equation}
\frac{n}{3m_{4}^{3}}\Bigg[2(\rho+\rho_{\phi})+3(p+p_{\phi})\Bigg]_{y=0}
\end{equation}
where energy density and pressure of non-minimally coupled scalar
field are given as follows
\begin{equation}
\rho_{\phi}=\left[\frac{1}{2}\dot{\phi}^{2}+n^{2}V(\phi)-6\alpha'H\dot{\phi}\right]_{y=0},
\end{equation}
\begin{equation}
p_{\phi}=\left[\frac{1}{2n^{2}}\dot{\phi}^{2}-V(\phi)+
\frac{2\alpha'}{n^2}\Big(\ddot{\phi}-\frac{\dot{n}}{n}\dot{\phi}\Big)+
4\alpha'\frac{H}{n^{2}}\dot{\phi}+\frac{2\alpha''}{n^2}\dot{\phi}^2
\right]_{y=0},
\end{equation}
and $H=\frac{\dot{a}}{a}$ is Hubble parameter. Note that part of the
effect of non-minimal coupling of the field $\phi$ is hidden in the
definition of the effective energy density and pressure which both
include non-minimal terms. Now using (26), since in the bulk
$G_{00}=0$, we find
\begin{equation}
\lim_{\epsilon\to+0}\bigg[\frac{n'}{n}\bigg]_{y=-\epsilon}^{y=+\epsilon}=
\bigg[\frac{\dot{a}'}{\dot{a}}\bigg]_{y=-\epsilon}^{y=+\epsilon}
\end{equation}
using relations (32) and (33) we find the following relation for
conservation of energy on the brane
\begin{equation}
\dot{\rho}+\dot{\rho}_{\phi}+3H\Big(\rho+\rho_{\phi}+p+p_{\phi}\Big)=6\alpha'\dot\phi
\Big(H^2+\frac{k}{a^2}\Big).
\end{equation}
Thus the non-minimal coupling of the scalar field to the Ricci
curvature on the brane through $\alpha(\phi)$ leads to the
non-conservation of the effective energy density.

To obtain the cosmological dynamics, we set $n(0,t)=1$. With this
gauge condition we recover usual time on the brane via
transformation $t=\int^{t}n(0,\eta)d\eta$. In this situation, our
basic dynamical variable is only $a(y,t)$ since $n(y,t)$ now is
given by
\begin{equation}
n(y,t)=\frac{\dot{a}(y,t)}{\dot{a}(0,t)}.
\end{equation}
where $H=\frac{\dot{a}(0,t)}{a(0,t)}$ is Hubble parameter on the
brane. Now we can write the basic set of cosmological equations for
a FRW brane in the presence of a non-minimally coupled scalar field.
The first of these equations is given by matching condition
\begin{equation}
\lim_{\epsilon\rightarrow+0}\left[\partial_{y}a\right]^{y=
+\epsilon}_{y=-\epsilon}(t)=\frac{m^{2}_{3}}{m^{3}_{4}}\Bigg[\alpha(\phi)
\bigg(\frac{\dot{a}^{2}}{n^{2}a}+ \frac{k}{a}\bigg)\Bigg]_{y=0}
-\Bigg[\frac{(\rho+\rho_{\phi})a}{3 m^{3}_{4}}\Bigg]_{y=0}.
\end{equation}
Insertion of\,\,$\frac{n'}{n}=\frac{\dot{a}'}{\dot{a}}$\,\, into
equations (24) and (27) yields
$$\frac{2}{3n^2}a'a^{3}G_{00}=\frac{\partial}{\partial
y}\bigg(\frac{\dot{a}^{2}}{n^2}a^{2}-a'^{2}a^{2}+ka^{2}\bigg)=0$$
and
$$\frac{2}{3}\dot{a}a^{3}G_{yy}=-\frac{\partial}{\partial t}\bigg(\frac{\dot{a}^{2}}{n^2}a^{2}-a'^{2}a^{2}+ka^{2}\bigg)=0.$$
These two equations imply that the Bin\'{e}truy {\it et al}\,\,[14]
integrals
\begin{equation}
{\cal
I}^{+}=\bigg[\Big(\frac{\dot{a}^{2}}{n^2}-a'^{2}+k\Big)a^{2}\bigg]_{y>0},
\end{equation}
and
\begin{equation}
{\cal
I}^{-}=\bigg[\Big(\frac{\dot{a}^{2}}{n^2}-a'^{2}+k\Big)a^{2}\bigg]_{y<0},
\end{equation}
are constant and if $a'$ is continuous on the brane then ${\cal
I}^{+}={\cal I}^{-}$. These equations along with scalar field
equation
\begin{equation}
\ddot{\phi}+\bigg(3\frac{\dot a}{a}-\frac{\dot
n}{n}\bigg)\dot{\phi}+n^{2}\frac{dV}{d\phi}-n^{2}\frac{d\alpha}{d\phi}R[q]=0,
\end{equation}
and
\begin{equation}
n(y,t)=\frac{\dot{a}(y,t)}{\dot{a}(0,t)}.
\end{equation}
constitute the basic dynamical equations of our model. In the
absence of transverse momentum, $\Upsilon_{0y}=0$, one can show that
${\cal I}^{+}={\cal I}^{-}$. In fact ${\cal I}^{\pm}$ can be
considered as initial conditions and these quantities reflect the
symmetry across the brane. In the case of ${\cal I}^{+}\neq{\cal
I}^{-}$ there can not be any symmetry across the brane. So we first
consider the case ${\cal I}^{+}={\cal I}^{-}$ in which follows. Our
cosmological equations on the brane now take the following
forms(note that $n(0,t)=1$)
\begin{equation}
\frac{\dot{a}^{2}(0,t)+k}{a^{2}(0,t)}=
\frac{(\rho+\rho_{\phi})}{3m^{2}_{3}\alpha(\phi)},
\end{equation}
\begin{equation}
\ddot{\phi}+3\frac{\dot{a}(0,t)}{a(0,t)}\dot{\phi}+\frac{dV(\phi)}{d\phi}
=\frac{d\alpha}{d\phi}R[q],
\end{equation}
\begin{equation}
{\cal I}=\Big[\dot{a}^{2}(0,t)-a'^{2}(y,t)+k\Big]a^{2}(y,t)
\end{equation}
\begin{equation}
n(y,t)=\frac{\dot{a}(y,t)}{\dot{a}(0,t)}.
\end{equation}
Using equation (46), the scale factor is calculated as follows
\begin{equation}
a^{2}(y,t)=a^{2}(0,t)+ \Big[\dot{a}^{2}(0,t)+k\Big]y^{2}
+2\bigg[\Big(\dot{a}^{2}(0,t)+k\Big) a^{2}(0,t)-{\cal
I}\bigg]^{\frac{1}{2}}y
\end{equation}
and therefore $n(y,t)$ is given by equation (47);
$$n(y,t)=\Bigg(a(0,t)+\ddot{a}(0,t)y^{2}+a(0,t)\frac{a(0,t)\ddot{a}(0,t)+
\dot{a}^{2}(0,t)+k}{\sqrt{\Big(\dot{a}^{2}(0,t)+k\Big)a^{2}(0,t)-{\cal
{I}}}}y\Bigg)$$
\begin{equation}
\times \Bigg[a^{2}(0,t)+ \Big[\dot{a}^{2}(0,t)+k\Big]y^{2}
+2\Big[\Big(\dot{a}^{2}(0,t)+k\Big) a^{2}(0,t)-{\cal
I}\Big]^{\frac{1}{2}}y\Bigg]^{\frac{-1}{2}}
\end{equation}
So, the components of 5-dimensional metric (22) are determined. If
we set initial conditions in such a way that ${\cal I}= 0$, we find
the following simple equations for cosmological dynamics
\begin{equation}
 a(y,t)=a(0,t)+\Big[\dot{a}^{2}(0,t)+k\Big]^{\frac{1}{2}}y,
\end{equation}
\begin{equation}
n(y,t)=1+\frac{\ddot{a}(0,t)}{\sqrt{\dot{a}^{2}(0,t)+k}}y.
\end{equation}
Therefore, our model allows for an embedding of the standard
Friedmann cosmology in the sense that the cosmological evolution of
the background metric on the brane can be described by the standard
Friedmann equation plus energy non-conservation on the brane.

So far we have discussed the case ${\cal I}^{+}={\cal I}^{-}$ with a
continuous warp factor across the brane. In the case of ${\cal
I}^{+}\neq{\cal I}^{-}$, there cannot be any symmetry across the
brane. In this case the basic set of dynamical equations is provided
by equations (39), (40) and (41) plus the non-conservation of the
effective energy density given by (37). In this case, evolution of
the scale factor on the brane is given by elimination of
$a'(y\longrightarrow\pm0,t)$ from the following generalized
Friedmann equation

$$\pm\bigg[\dot{a}^{2}(0,t)+k-a^{-2}(0,t){\cal
I}^{+}\bigg]^{\frac{1}{2}}\mp\bigg[\dot{a}^{2}(0,t)+k-a^{-2}(0,t){\cal
I}^{-}\bigg]^{\frac{1}{2}}$$
\begin{equation}
\quad\quad\quad\quad\quad\quad=\alpha(\phi)\frac{m_{3}^{2}}{m_{4}^{3}}
\bigg(\frac{\dot{a}^{2}(0,t)+k}{a(0,t)}\bigg)-
\frac{(\rho+\rho_{\phi})a(0,t)}{3m^{3}_{4}}.
\end{equation}
This is the most general form of the modified Friedmann equation for
our non-minimal framework. After determination of $a(0,t)$, since
${\cal I}^{\pm}$ are constants, \, $a(y,t)$ can be calculated from
(46). This is the full dynamics of the system. Note that in the case
where the right hand side of equation (52) is negative, at least one
sign in the left hand side should be negative depending on initial
conditions. However, the dynamics of the problem does not require
symmetry across the brane. Therefore, we have shown the possibility
of embedding of FRW cosmology in DGP scenario with a 4D
non-minimally coupled scalar field on the brane and equation (52) is
the most general form of FRW equation in this embedding.

\section{Dark Radiation Formalism with Non-Minimal Coupling}
In order to show that our model is consistent with dark energy
formulation of Friedmann equation, we first obtain non-minimal
extension of this formalism and then we show that this equation can
be obtained in our framework more easily. The effective Einstein
equation on the brane is given by[24]
\begin{equation}
G_{\mu\nu}=\frac{\Pi_{\mu\nu}}{m_{4}^{6}}-{\cal{E}}_{\mu\nu},
\end{equation}
where
\begin{equation}
\Pi_{\mu\nu}=-\frac{1}{4}{\cal{T}}_{\mu\sigma}{{\cal{T}}_{\nu}}^{\sigma}+
\frac{1}{12}{\cal{T}}{\cal{T}}_{\mu\nu}+
\frac{1}{8}g_{\mu\nu}\Big({\cal{T}}_{\rho\sigma}{\cal{T}}^{\rho\sigma}-\frac{1}{3}{\cal{T}}^{2}\Big),
\end{equation}
and  ${\cal{T}}_{\mu\nu}$ is given by equation (7). Also we have
\begin{equation}
{\cal{E}}_{\mu\nu}=C_{MRNS}\,\, n^{M}\,\,n^{N}
{g^{R}}_{\mu}\,\,{g^{S}}_{\nu}
\end{equation}
where $C_{MRNS}$ is five dimensional Weyl tensor and $n_{A}$ is the
spacelike unit vector normal to the brane. Now, using equation (53)
we find
\begin{equation}
{G^{0}}_{0}=\frac{{\Pi^{0}}_{0}}{m_{4}^{6}}-{{\cal{E}}^{0}}_{0}
\end{equation}
where for FRW universe we have
\begin{equation}
{G^{0}}_{0}=-3\Big(H^{2}+\frac{k}{a^{2}}\Big).
\end{equation}
Similarly, for space components we have
\begin{equation}
{G^{i}}_{j}=\frac{{\Pi^{i}}_{j}}{m_{4}^{6}}-{{\cal{E}}^{i}}_{j}
\end{equation}
where
\begin{equation}
{G^{i}}_{j}=-\Big(2\dot{H}+3H^{2}+\frac{k}{a^{2}}\Big){\delta^{i}}_{j}.
\end{equation}
Now using equation (54) we find \,
${\Pi^{0}}_{0}=-\frac{1}{12}\Big({{\cal{T}}^{0}}_{0}\Big)^{2}$ \,
and\,
${\Pi^{i}}_{j}=-\frac{1}{12}{{\cal{T}}^{0}}_{0}\Big({{\cal{T}}^{0}}_{0}-
2{{\cal{T}}^{1}}_{1}\Big){\delta^{i}}_{j}$. Also, equation (7) gives
\begin{equation}
{{\cal{T}}^{0}}_{0}=-(\rho+\rho_{\phi})-m_{3}^{2}\alpha(\phi){G^{0}}_{0}
\end{equation}
and
\begin{equation}
{{\cal{T}}^{i}}_{j}=-(p+p_{\phi})\delta^{i}_{j}-m_{3}^{2}\alpha(\phi){G^{i}}_{j},
\end{equation}
where $\rho_{\phi}$ and $p_{\phi}$ are given by (34) and (35) with
$n(0,t)=1$. These equations lead us to the following generalized
Friedmann equation
\begin{equation}
3\bigg(H^{2}+\frac{k}{a^{2}}\bigg)={{\cal{E}}^{0}}_{0}+
\frac{1}{12m_{4}^{6}}\bigg[\rho+\rho_{\phi}-3m_{3}^{2}\alpha(\phi)\Big(H^{2}+\frac{k}{a^{2}}\Big)\bigg]^{2}.
\end{equation}
Using Codazzi equation we have
$\nabla^{\nu}{\cal{E}}_{\mu\nu}=0$\,(see for example [24,31]).
Therefore we find
$\dot{{{\cal{E}}}^{0}}_{0}+4H{{\cal{E}}^{0}}_{0}=0$ \,\, which
integration gives
${{\cal{E}}^{0}}_{0}=\frac{{{\cal{E}}_{0}}}{a^{4}}$ with
${\cal{E}}_{0}$ as an integration constant.  Therefore, equation
(62) can be re-written as follows
\begin{equation}
H^{2}+\frac{k}{a^2}=\frac{1}{3m_{3}^{2}\alpha(\phi)}
\bigg(\rho+\rho_{\phi}+\rho_{0}\bigg[1+\varepsilon
\sqrt{1+\frac{2}{\rho_{0}}
\Big[\rho+\rho_{\phi}-m_{3}^{2}\alpha(\phi)
\frac{{{\cal{E}}_{0}}}{a^{4}}\Big]}\,\,\bigg]\,\,\bigg).
\end{equation}
where $\rho_{0}=\frac{6m_{4}^{6}}{m_{3}^{2}\alpha(\phi)}$ \, and \,
$\varepsilon=\pm1$ \, shows the possibility of existence of two
different branches of FRW equation. In the high energy regime where
$\frac{\rho+\rho_{\phi}}{\rho_{0}}\gg 1$, we find
\begin{equation}
H^{2}\approx\frac{1}{3m_{3}^{2}\alpha(\phi)}\Big(\rho+\rho_{\phi}+\varepsilon\sqrt{2(\rho+\rho_{\phi})\rho_{0}}\,\,\Big)
\end{equation}
which describes a four dimensional gravity with a small correction.
In the low energy regime where $\frac{\rho+\rho_{\phi}}{\rho_{0}}\ll
1$, we find
\begin{equation}
H^{2}\approx\frac{1}{3m_{3}^{2}\alpha(\phi)}\bigg[(1+\varepsilon)(\rho+\rho_{\phi})+(1+\varepsilon)\rho_{0}
-\frac{\varepsilon}{4}\frac{(\rho+\rho_{\phi})^{2}}{\rho_{0}}\bigg].
\end{equation}
Now we show that this result can be obtained in our formalism more
easily. For this purpose we show that relation (52) for the case
with ${\cal I}^{+}={\cal I}^{-}\equiv{\cal I}$ and a discontinuous
warp factor across the $Z_{2}$ symmetric brane leads to this result
with some simple algebra. For simplicity, we define $ x\equiv
H^{2}+\frac{k}{a^{2}},$\,\, $b\equiv\rho+\rho_{\phi},$\,\,
$y\equiv\alpha(\phi)m_{3}^{2},$\,\, and \, $z\equiv m_{4}^{3}$.
With these definitions, equation (52) (with upper sign for
instance), transforms to the following form
\begin{equation}
\bigg(x-\frac{{\cal
I}^{+}}{a^4}\bigg)^{\frac{1}{2}}+\bigg(x-\frac{{\cal
I}^{-}}{a^4}\bigg)^{\frac{1}{2}}=\frac{y}{z}x-\frac{b}{3z}.
\end{equation}
Solving this equation for $x$ (with ${\cal I}^{+}={\cal
I}^{-}\equiv{\cal I}$) gives the following result
\begin{equation}
x=\frac{\frac{by}{3z^2}+2\pm\sqrt{\Big(\frac{by}{3z^2}+2\Big)^{2}
-\frac{y^2}{z^{2}}\Big(\frac{b^2}{9z^2}+\frac{4{\cal
I}}{a^4}\Big)}}{\frac{y^2}{z^2}}.
\end{equation}
A little algebraic manipulation gives
\begin{equation}
x=\frac{1}{3y}\Bigg[b+\frac{6z^2}{y}\pm\frac{6z^2}{y}
\sqrt{1+\frac{by}{3z^2}-\frac{{\cal I}y^2}{a^{4}z^{2}}}\Bigg].
\end{equation}
Considering both plus and minus signs in equation (52) and using
original quantities we obtain
\begin{equation}
H^{2}+\frac{k}{a^2}=\frac{1}{3m_{3}^{2}\alpha(\phi)}\bigg(\rho+\rho_{\phi}+
\rho_{0}\bigg[1+\varepsilon
\sqrt{1+\frac{2}{\rho_{0}}\Big[\rho+\rho_{\phi}-m_{3}^{2}
\alpha(\phi)\frac{{{\cal{E}}_{0}}}{a^{4}}\Big]}\bigg]\,\,\bigg).
\end{equation}
where
$\rho_{0}\equiv\frac{6z^2}{y}=\frac{6m_{4}^{6}}{m_{3}^{2}\alpha(\phi)}$
\, and \, ${\cal{E}}_{0}=3{\cal I}$\,\, is a constant. This analysis
shows the consistency of our formalism with dark-radiation formalism
presented above.

\section{Cosmological Considerations}
Now to discover cosmological implications of non-minimally coupled
scalar field on DGP braneworld, we proceed as follows: the evolution
of scalar field on the brane for spatially flat FRW metric ($k=0$)
is given by the following equations
\begin{equation}
H_{0}^{2}=\frac{1}{3\alpha(\phi) m^{2}_{3}}(\rho+\rho_{\phi}),
\end{equation}
and
\begin{equation}
\ddot{\phi}+3H_{0}\dot{\phi}+\frac{dV(\phi)}{d\phi}=\alpha'R[q],
\end{equation}
where $H_{0}\equiv\frac{\dot{a}(0,t)}{a(0,t)}$ and for $n(0,t)=1$
(on the brane) we have
\begin{equation}
\rho_{\phi}=\frac{1}{2}\dot{\phi}^{2}+V(\phi)-6\alpha'H\dot{\phi}.
\end{equation}
Ricci scalar on the brane is given by the following relation
\begin{equation}
R=6\frac{\ddot{a}}{a}+6\Big(\frac{\dot{a}}{a}\Big)^{2}.
\end{equation}
In the absence of ordinary matter on the brane we set $\rho=0$. To
find cosmological implications of our model we should solve
equations (70) and (71) using (72) and (73). For comparison purpose,
we first give an overview of the minimal case, that is, the case
with \, $\alpha=constant$ which has been discussed with details in
reference [30]. First we should specify the form of the scalar field
potential $V(\phi)$. We use the following potential which has been
motivated from theories of modified gravity with Lagrangian of the
type ${\cal L}(R)=R-\frac{\mu^{4}}{R}$\, with\, $R^{-1}$ term[11,30]
\begin{equation}
V(\phi)\simeq\mu^{2}m^{2}_{3}\exp\left(-\sqrt{\frac{3}{2}}\frac{\phi}{m_{3}}\right).
\end{equation}
In this case we find \,\, $ a(0,t)\propto t^{4/3}$ \,\,and \,\,$
\phi\propto-\frac{4}{3}\ln t$\,\, for the evolution of scale factor
and scalar field on the brane respectively. Obviously, this predicts
a power-law acceleration on the brane. This result is consistent
with the observational results similar to quintessence with the
equation of state parameter $-1<w_{DE}<-\frac{1}{3}$, [6,20]. In
this case, using (48) and (49), the evolution of $a(y,t)$ and
$n(y,t)$ is given by the following equations [30]
\begin{equation}
a^{2}(y,t)=C^{2}\bigg(t^{\frac{8}{3}}+\frac{16}{9}t^{\frac{2}{3}}y^{2}\bigg)+
2\bigg(\frac{16}{9}C^{4}t^{\frac{10}{3}}-{\cal
I}\bigg)^{\frac{1}{2}}y
\end{equation}
and
\begin{equation}
n(y,t)=C\bigg[t^{\frac{4}{3}}+\frac{4}{9}t^{-\frac{2}{3}}y^{2}
+\frac{20}{9}C^{2} t^{\frac{8}{3}}
\bigg(\frac{16}{9}C^{4}t^{\frac{10}{3}}-{\cal
I}\bigg)^{\frac{-1}{2}}y\bigg]\frac{1}{a(y,t)},
\end{equation}
where $C$ is a constant and ${\cal
I}=\Big[\dot{a}^{2}(0,t)-a'^{2}(y,t)+k\Big]a^{2}$. If we set the
initial conditions in such a way that ${\cal I}=0$, these results
become very simple
\begin{equation}
a(y,t)=C\Big(t^{\frac{4}{3}}+\frac{4}{3}t^{\frac{1}{3}}y\Big),\quad\quad\quad\quad\quad
n(y,t)=C\Big(1+\frac{y}{3t}\Big).
\end{equation}
In the case where $\alpha(\phi)$ is not just a constant, the
situation becomes more complicated since now the kind of the
cosmological solutions depend explicitly on the value of non-minimal
coupling. We first try to obtain a necessary condition for the
acceleration of the universe in an specific model. Suppose that
$\rho=0$. By definition, the required condition for acceleration of
the universe is \,$\rho_{\phi}+3p_{\phi}<0$. Using equations
defining $\rho_{\phi}$ and $p_{\phi}$, we find
\begin{equation}
(1+3\alpha'')\dot{\phi}^{2}-V(\phi)+3\alpha'(H\dot{\phi}+\ddot{\phi})<0.
\end{equation}
Using Klein-Gordon equation (71), this relation can be rewritten as
follows
\begin{equation}
(1+3\alpha'')\dot{\phi}^{2}-V(\phi)+3\alpha'^{2} R-
6\alpha'H_{0}\dot{\phi}-3\alpha'\frac{dV}{d\phi}<0.
\end{equation}
Finally, using (72), this can be written as
\begin{equation}
\rho_{\phi}-2V(\phi)+(\frac{1}{2}+3\alpha'')\dot{\phi}^{2}+3\alpha'^{2}
R-3\alpha'\frac{dV}{d\phi}<0.
\end{equation}
This is a general condition to have an accelerating universe with
non-minimally coupled scalar field on the brane [22,35]. To proceed
further, we assume weak energy condition $\rho_{\phi}\geq 0$.
Motivated from several theoretical evidences( for example: conformal
coupling in general relativity and other metric theories[35],
renormalization group study of $\lambda\phi^{4}$ theory[36] and
large $N$ limit of the Nambu-Jona-Lasinio model[37], (see also [22]
and references therein), in which follows we set
$\alpha(\phi)=\frac{1}{2}(1-\xi\phi^{2})$  with
$\xi\leq\frac{1}{6}$. Therefore we find
\begin{equation}
V-\frac{3\xi}{2}\phi\frac{dV}{d\phi}>0.
\end{equation}
As an example, suppose that $V(\phi)=\lambda \phi^{n}$. In this case
with \, $\xi\leq\frac{1}{6}$\,  and using (81), we find
\begin{equation}
\lambda\bigg(1-\frac{3n\xi}{2}\bigg)>0.
\end{equation}
If we assume a positive scalar field potential with $\lambda >0$,
the condition for accelerating expansion restricts $\xi$ to the
values which $\xi\leq\frac{2}{3n}$.\, So in the presence of
non-minimally coupled scalar field, it is harder to achieve
accelerating universe with usual potentials.

Now with above definitions of non-minimal coupling and scalar field
potential, equations (70) and (71) take the following forms
\begin{equation}
\frac{\dot{a}^2}{a^2}=\frac{2}{3}\big(1-\xi\phi^{2}\big)^{-1}\bigg[\frac{1}{2}\dot{\phi}^{2}+\lambda\phi^{n}+6\xi\phi\frac{\dot{
a}}{a}\dot{\phi}\bigg],
\end{equation}
and
\begin{equation}
\ddot{\phi}+ 3\frac{\dot{
a}}{a}\dot{\phi}+n\lambda\phi^{n-1}+6\xi\phi\Big[\frac{\ddot{
a}}{a}+\frac{\dot{a}^2}{a^2}\Big]=0.
\end{equation}
In order to study late time behavior of these equations, we try the
following ansatz,
\begin{equation}
a(t)\approx A \,t^{\nu},\quad\quad \phi(t)\approx B\, t^{-\mu}
\end{equation}
where we have assumed a decreasing power law ansatz for scalar
field. With these choices and setting $n=4$, equation (83) gives
\begin{equation}
\frac{3}{2}\frac{\nu^2}{t^2}-\frac{3}{2}\xi
B^{2}\nu^{2}t^{-2\mu-2}=\Big(\frac{1}{2}B^{2}\mu^{2}-6\xi
B^{2}\mu\nu\Big)t^{-2\mu-2}+\lambda B^{4} t^{-4\mu}
\end{equation}
On the other hand, equation (84) gives
\begin{equation}
\bigg[\mu(\mu+1)-3\mu\nu+6\xi(2\nu^{2}+\nu)\bigg]t^{-\mu-2}+4\lambda
B^{3} t^{-3\mu}=0.
\end{equation}
Considering terms of order ${\cal{O}}(t^{-\mu-2})$, equation (86)
gives $\xi\geq\frac{1}{12}$ if we require a positive and real $\nu$.
So for this special ansatz $\xi$ is restricted to the condition
$\frac{1}{12}\leq\xi\leq\frac{1}{6}$.\, With the same procedure,
equation (87) gives
\begin{equation}
\mu^{2}+(1-3\nu)\mu+12\xi\nu^{2}+6\xi\nu=0.
\end{equation}
Again, positivity and reality of solutions for $\mu$ lead to the
following constraint
\begin{equation}
(9-48\xi)\nu^{2}-(6+24\xi)\nu+1\geq 0
\end{equation}
where for $\xi=\frac{1}{12}$ and taking equality, we find
$\nu=\frac{4\pm\sqrt{11}}{5}$ which plus sign obviously leads to an
power-law accelerated expansion. So, we conclude that although in
the presence of non-minimally coupled scalar field, accelerated
expansion of universe is harder to achieve relative to minimally
coupled scalar field case, but with a suitable choice of nonminimal
coupling it is possible to explain this accelerated expansion. \\
On the other hand, with a nonminimal coupling of the kind
$-\frac{1}{2}\xi R \phi^2$, we can define
\begin{equation}
G_{eff}=\frac{G}{1-\frac{\phi^2}{\phi_{c}^{2}}},
\end{equation}
where $\phi_{c}^{2}\equiv\frac{m_{3}^{2}}{8\pi \xi}$. In order to
connect to our present universe, $G_{eff}$ needs to be positive for
the case of the positive $\xi$, which yields $|\phi|<
\phi_{c}=\frac{m_{3}}{\sqrt{8\pi \xi}}$. When $\xi$ is negative,
such a constraint is absent. By performing a conformal
transformation[40] to transform to the Einstein frame, we define
\begin{equation}
\hat{V} \equiv \frac{V(\phi)}{(1-\xi \kappa^{2}\phi^{2})^{2}}
\end{equation}
Figure $3$ shows the behavior of this effective potential with
different values of nonminimal coupling and $V(\phi)$ defined in
(74). As we see, when $\xi$ is positive, since the potential
$\hat{V}(\phi)$ is flat in the region of $\phi>0$ compared with the
$\xi=0$ case, we can expect assisted inflation (due to NMC) to occur
in this region. When $\xi$ is negative, assisted inflation can be
realized in the region of $\phi<0$.\\

\begin{figure}
\includegraphics{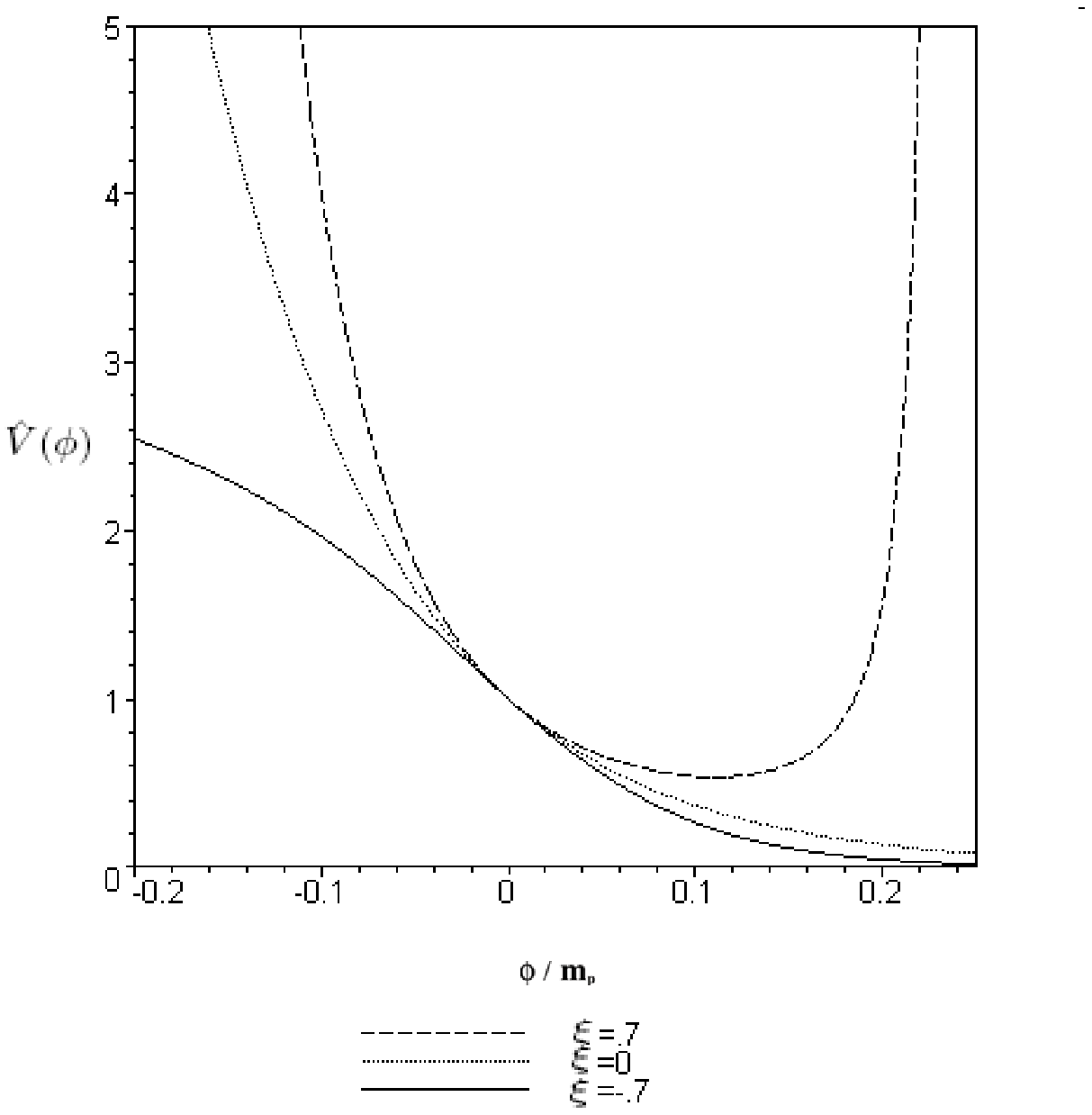} \vspace{7cm} \caption{\small { An
effective potential $\hat{V}(\phi)$ in the Einstein frame in the
case of $\xi=0.7, 0, -0.7$ with $p=1/2$ (top, middle, bottom). When
$\xi$ is positive, since the potential $\hat{V}(\phi)$ is flat in
the region of $\phi>0$ compared with the $\xi=0$ case, we can expect
assisted inflation to occur in this region. When $\xi$ is negative,
assisted inflation can be realized in the region of $\phi<0$.}}
\label{fig:1}
\end{figure}

Based on some holographic dark energy model, the value of NMC,
$\xi$ to have an accelerated universe is restricted to the interval
$0.146\preceq \xi \preceq 0.167$\,[44]. On the other hand, current
experimental limits on the time variation of $G$ constraint the
nonminimal coupling as $ - 10^{-2}\preceq \xi \preceq
10^{-2}$\,[45]. Solar system experiments such as Shapiro time delay
and deflection of light[33] have constraint Brans-Dike parameter to
be $\omega_{BD}>500$. This constraint leads to the result of
$|\xi|\preceq 2.2\times10^{-2}$ for non-minimal coupling [45].

\section{Quintessence Model}
To study quintessence model within our setting, we write the
equation of state for scalar field which takes the following form
\begin{equation}
w\equiv\frac{p_{\phi}}{\rho_{\phi}}=\frac{\dot{\phi}^{2}-2V(\phi)+
4\Big(\alpha'\ddot{\phi}+2H\alpha'\dot{\phi}+\alpha''\dot{\phi}^2\Big)}{\dot{\phi}^{2}+2V(\phi)-12\alpha'H\dot{\phi}}.
\end{equation}
Causality implies that $|w|\leq 1$. When $\dot{\phi}=0$, we obtain
$p_{\phi}=-\rho_{\phi}$. In this case $\rho_{\phi}$ is independent
of $a$ and $V(\phi)$ plays the role of a cosmological constant. If
$V(\phi)=0$, we find
\begin{equation}
w=\frac{\dot{\phi}^{2}+
4\Big(\alpha'\ddot{\phi}+2H\alpha'\dot{\phi}+\alpha''\dot{\phi}^2\Big)}{\dot{\phi}^{2}-12\alpha'H\dot{\phi}}.
\end{equation}
In the case of minimal coupling ($\alpha=const.$) this is
corresponding to a massless scalar field which plays the role of a
{\it stiff matter} since $\rho_{\phi}\sim\frac{1}{a^{6}}$. However,
in our case with non-minimally coupled scalar field, the situation
is very different since now $\alpha$ plays a crucial role and
depending on the form of $\alpha$ we may obtain some traces of stiff
matter or fail to have such extreme case. So, in principle our model
provide a mechanism to avoid stiff matter. In the minimal case when
$\dot{\phi}^{2}< V(\phi)$, we obtain
$p_{\phi}<-\frac{\rho_{\phi}}{3}$ which shows a late-time
accelerating universe. In this case, since
$V(\phi)=\frac{1}{2}(1-w)\rho_{\phi}$\,
and\,\,$\phi=\sqrt{3(1+w)}\ln a$,\, the following potential which is
a Liouville-type potential, decreases when scalar field $\phi$
increases and therefore gives the required quintessence
\begin{equation}
V(\phi)=V_{0}\exp{\Big(-\sqrt{3(1+w)}\phi}\Big),
\end{equation}
where $0<3(1+w)<2$. On the other hand, for non-minimal case we find
\begin{equation}
V(\phi)=\frac{1}{2}\alpha(1-w)\rho+
\Big(\alpha'\ddot{\phi}+5H\alpha'\dot{\phi}+\alpha''\dot{\phi}^2\Big).
\end{equation}
In this case the quintessence potential has an explicit dependence
on the non-minimal coupling and its derivatives with respect to
$\phi$. To have a quintessence model, we should impose some limits
on the shape of this non-minimal coupling. Since based on
quintessence proposal the dark energy of the universe is dominated
by the potential of scalar field, we should impose suitable
constraints on non-minimal coupling such that scalar field potential
decrease when scalar field increases. We assume ordinary matter on
the brane has energy density $\rho\propto a^{-3}$ and vanishing
pressure $p=0$. Now total energy density becomes $\rho_{T}\equiv
\rho_{\phi}+\rho$. The necessary and sufficient condition for
acceleration of the universe, $\rho_{T}+3P_{T}<0$, leads to the
following relation
\begin{equation}
(1+3\alpha'')\dot{\phi}^{2}-V(\phi)+3\alpha'^{2} R-
6\alpha'H_{0}\dot{\phi}-3\alpha'\frac{dV}{d\phi}+\frac{\rho}{2}<0,
\end{equation}
which can be written as
\begin{equation}
\rho_{\phi}-2V(\phi)+(\frac{1}{2}+3\alpha'')\dot{\phi}^{2}+3\alpha'^{2}
R-3\alpha'\frac{dV}{d\phi}+\frac{\rho}{2}<0.
\end{equation}
This is a general constraint to have a quintessence scenario in the
presence of non-minimal coupling. If we assume that $\rho$ and
$\rho_{\phi}$ are non-negative, with
$\alpha(\phi)=\frac{1}{2}(1-\xi\phi^{2})$ and  $\xi\leq\frac{1}{6}$,
we find
\begin{equation}
V-\frac{3\xi}{2}\phi\frac{dV}{d\phi}>0.
\end{equation}
To have quintessential expansion, this constraint with
$0<\xi\leq\frac{1}{6}$, restrict the form of scalar field potential
to potentials $V(\phi)>0$ where
$\frac{d}{d\phi}\bigg[\ln\Big[\frac{V}{V_{0}}\Big(\frac{\phi_{0}}{\phi}\Big)^{\omega}
\exp\Big(-\frac{\phi^2}{6}\Big)\Big]\bigg]<0$ where
$\omega=\frac{1}{3\xi}\Big(1-\frac{\Omega}{2\Omega_{\phi}}|_{t=t_{0}}\Big)$,\,\,
$\Omega=\rho/\rho_{c}$,\, $\Omega_{\phi}=\rho_{\phi}/\rho_{c}$\, and
$\frac{\Omega}{\Omega_{\phi}}$ has been approximated by its present
value[22]. So, to have a quintessential expansion with non-minimal
coupling with $0<\xi\leq 1/6$\, we need a potential that does not
grow faster than\,
$f(\phi)=V_{0}\Big(\frac{\phi_{0}}{\phi}\Big)^{\omega}\exp\big(\frac{\phi^2}{6}\big)$\,
with variation of \,$\phi$. An inverse-power-law potential such as
$V(\phi)=\mu^{4+\delta}\phi^{-\delta}$, an exponential potential
such as $V(\phi)=\mu^{4}\exp{(-\lambda\frac{\phi}{m_{4}})}$\ and
several other possibilities[38,39] provide suitable potentials for
quintessence. But obviously quintessence model in the presence of
non-minimal coupling needs more artificial arguments than minimal
case.

\section{Summary and Conclusions}
In this paper we have considered the DGP model with a non-minimally
coupled scalar field on the brane. As we have explained, the
introduction of non-minimal coupling is not just a matter of taste;
it is forced upon us in many situations of physical and cosmological
interests such as quantum corrections to the scalar field theory and
its renormalizability in curved spacetime. In the spirit of DGP
inspired gravity, we have studied the effect of an induced gravity
term which is an arbitrary function of a scalar field on the brane.
We have presented four-dimensional equations on a DGP brane with a
scalar field non-minimally coupled to the induced Ricci curvature,
embedded in a five-dimensional Minkowski bulk. This is an extension
to a braneworld context of scalar-tensor (Brans-Dicke) gravity.
Cosmological implications of both minimal and non-minimal extension
of our model are studied. In minimal case, we have considered an
exponentially decreasing potential which has been motivated by
modified theory of gravity with $R^{-1}$ modification. In the
non-minimal case however, we have considered potential of the type
$V(\phi)=\lambda{\phi^n}$. We have studied the weak field limit of
our model and it has been shown that the mass density of ordinary
matter on the brane should be modified by the addition of the
effective mass density attributed to the non-minimally coupled
scalar field on the brane. Also in this case crossover scale of DGP
scenario is modified by the presence of non-minimal coupling. We
have discussed the role of nonminimal coupling in crossover
distance. Considering the case of FRW brane, we have obtained the
evolution of the metric and scalar field by solving the field
equations in the limit of small curvature. Our solutions for minimal
case predict a power-law acceleration on the brane supporting
observed late-time acceleration. For non-minimal case we have shown
that by a suitable choice of non-minimal coupling and scalar field
potential one can achieve accelerated expansion in some special
cases. However, As Faraoni has shown, in the presence of
non-minimally coupled scalar field accelerated expansion of universe
is harder to achieve relative to minimally coupled scalar field
case. We have studied quintessence model in our framework and it has
been shown that for a restrict class of non-minimal coupling one can
achieve quintessence potential. As an important achievement our
analysis shows that DGP model allows for an embedding of the
standard Friedmann cosmology in the sense that the cosmological
evolution of the background metric on the brane can entirely be
described by the standard Friedmann equation plus energy
non-conservation on the brane. As we have shown, our model gives
dark energy extension of Friedmann equation more easily than
standard framework. Some observational and experimental constraint
on nonminimal coupling are discussed. The issue of Non-minimal
inflation on warped DGP braneworld and confrontation with WMAP3 data
will be reported in a separated paper[46].\\

{\bf Acknowledgement}\\
It is a pleasure to appreciate members of the Centre for Particle
Theory at Durham University, specially Professor Ruth Gregory for
hospitality during my sabbatical leave. Also I would like to
appreciate referee for his/her important contribution in this work.

\end{document}